\documentstyle[aaspp4,psfig]{article}

\def\etal{{\sl et al.}}
\def\lsim{\hbox{ \rlap{\raise 0.425ex\hbox{$<$}}\lower 0.65ex\hbox{$\sim$} }}
\def\gsim{\hbox{ \rlap{\raise 0.425ex\hbox{$>$}}\lower 0.65ex\hbox{$\sim$} }}
\def\arcmin{\hbox{$^\prime$}}
\def\arcsec{\hbox{$^{\prime\prime}$}}
\def\fd{\hbox{$~\!\!^{\rm d}$}}
\def\fh{\hbox{$~\!\!^{\rm h}$}}
\def\fm{\hbox{$~\!\!^{\rm m}$}}
\def\fs{\hbox{$~\!\!^{\rm s}$}}
\def\ale{\mathrel{\hbox{\rlap{\hbox{\lower4pt\hbox{$\sim$}}}\hbox{$<$}}}}
\def\age{\mathrel{\hbox{\rlap{\hbox{\lower4pt\hbox{$\sim$}}}\hbox{$>$}}}}

\slugcomment{Submitted to The Astrophysical Journal (Letters)}

\lefthead{Bloom \etal~  }
\righthead{The Host Galaxy of GRB 990123}

\begin{document}

\title{The Host Galaxy of GRB 990123}

\author
{J. S. Bloom$^1$, S. C. Odewahn$^1$, S. G. Djorgovski$^1$, S. R. Kulkarni$^1$,}

\author {F. A. Harrison$^1$, C. Koresko$^1$, G.~Neugebauer$^1$,
L.~Armus$^2$, D. A. Frail$^2$, R. R. Gal$^1$,}

\author
{R. Sari$^1$, G. Squires$^1$, G. Illingworth$^3$, D. Kelson$^4$, F. Chaffee$^5$, }

\author
{R. Goodrich$^5$, M. Feroci$^6$, E. Costa$^6$, L. Piro$^6$, F. Frontera$^7$,}

\author
{S. Mao$^8$, C. Akerlof$^9$, T. A. McKay$^9$}

\bigskip 

\affil{$^1$ Palomar Observatory 105--24, California Institute of Technology,
            Pasadena, CA 91125, USA}

\affil{$^2$ Infrared Processing \&\ Analysis Center, Caltech, 
  Pasadena, CA 91125, USA}

\affil{$^3$ National Radio Astronomy Observatory, Socorro, NM 87801, USA}

\affil{$^4$ Univ.~of California Observatories/Lick Observatory, University of
            California at Santa Cruz, Santa Cruz, CA 95064, USA} 

\affil{$^5$ Department of Terrestrial Magnetism, Carnegie Inst. of Washington,
            Washington, DC 20015}

\affil{$^6$ W.~M.~Keck Observatory, 65-0120 Mamalahoa Highway, Kamuela, Hawaii
            96743, USA} 

\affil{$^7$ Instituto di Astrofisica Spaziale, CNR, via Fosso del Cavaliere, 
            Roma, I-00133, Italy}

\affil{$^8$ ITESRE-CNR, Via Gobetti 101, I-40129, Bologna, Italy}

\affil{$^9$ Max-Planck-Institut f\"ur Astrophysik, Karl-Schwarzschild-Strasse 1, 
            85740, Garching, Germany}

\affil{$^{10}$ Department of Physics, Univ. of Michagan, Ann Arbor, MI 48109, USA}

\begin{abstract}

We present deep images of the field of $\gamma$-ray burst (GRB) 990123
obtained in a broad-band UV/visible bandpass with the Hubble Space
Telescope, and deep near-infrared images obtained with the Keck-I 10-m
telescope.  Both the HST and Keck images show that the optical
transient (OT) is clearly offset by 0.6 arcsec from an extended
object, presumably the host galaxy.  This galaxy is the most likely
source of the metallic-line absorption at $z = 1.6004$ seen in the
spectrum of the OT.  With magnitudes $V_{C} \approx 24.6 \pm 0.2$ and $K =
21.65 \pm 0.30$ mag this corresponds to an $L \sim 0.7 ~L_*$ galaxy,
assuming that it is located at $z = 1.6$.  The estimated unobscured
star formation rate is $SFR \sim 6 ~M_\odot$ yr$^{-1}$, which is not
unusually high for normal galaxies at comparable redshifts.  The
strength of the observed metallic absorption lines is suggestive of a
relatively high metallicity of the gas, and thus of a chemically
evolved system which may be associated with a massive galaxy.  It is
also indicative of a high column density of the gas, typical of damped
Ly$\alpha$ systems at high redshifts.  We conclude that this is the
host galaxy of GRB 990123. No other obvious galaxies are detected
within the same projected radius from the OT.  There is thus no
evidence for strong gravitational lensing magnification of this burst,
and some alternative explanation for its remarkable energetics may be
required.  The observed offset of the OT from the center of its
apparent host galaxy, $5.5 \pm 0.9$ proper kpc (projected) in the
galaxy's rest-frame, both refutes the possibility that GRBs are related
to galactic nuclear activity and supports models of GRBs which involve
the death and/or merger of massive stars.  Further, the HST image
suggests an intimate connection of GRB 990123 and a star-forming
region.

\end{abstract}


\section{Introduction}

A great deal of progress has been achieved in our understanding of
cosmic $\gamma$--ray bursts (GRBs) over the past two years.  The
breakthrough development was the precise localization of bursts by the
BeppoSAX satellite (\cite{boe97}), which led to the discovery of
long-lived afterglows of GRBs, ranging from x-rays (\cite{cos97}), to
optical (\cite{jvan97}) and radio (\cite{fra97}).  This, in turn, has
opened the possibility of detailed physical studies of the afterglows,
and measurements of their distances.

To date, several optical transients (OTs) associated with GRBs have been found,
and in almost every case a faint galaxy was found at the same location (to 
within a fraction of an arcsecond) after the OT has faded.  So far, redshifts 
have been obtained for four such GRB host galaxies: 
$z = 0.835$ for GRB 970508 (\cite{met97}, \cite{bloom98a}),
$z = 3.428$ for GRB 971214 (\cite{kul98}),
$z = 0.966$ for GRB 980703 (\cite{djo98}), and
$z = 1.0964$ for GRB 980613 (\cite{djo99a}).
These measurements have established that most or all GRBs are located at 
cosmological distances (\cite{pac95}), involving substantial energy release
(isotropic equivalent $E \gsim 10^{52 \pm 1}$ in the $\gamma$-rays alone).  

While the ultimate origin of GRBs is still not established, studies of
their afterglows provide several crucial constraints for theoretical
models.  First, the measurement of distances establishes the
energetics of the bursts, modulo the unknown beaming factor.  Second,
detailed studies of the afterglow light curves over a range of
wavelengths can constrain the physical parameters of the afterglows,
including the energetics and beaming.  Finally, the location of the
afterglows within their host galaxies and measurements of the star
formation rates (SFR) in these galaxies can constrain the nature of
the population of GRB progenitors.

The two leading models for GRBs involve the formation of a black hole
(BH): either via coalescence of a massive stellar remnant binary
(eg.~BH--NS, NS--NS; \cite{pac86},\cite{Goodman86},\cite{Nar92}) or
direct collapse of a massive star (\cite{Woos93}, \cite{pac98}).  Both
models predict that GRBs rates should strongly correlate with the
cosmic star-formation rates (SFR) and so most GRBs should occur in the
redshift range $z = 1 - 2$.  The former model predicts a tight
spatial correlation between GRBs and star-forming regions in the disk.
In the latter scenario, however, the coalescence site of a NS--NS
binary can be quite distant ($\age$ few kpc) from the stellar birth
site (see \cite{bloom98b}).  GRBs could also be associated with
nuclear black holes (AGN); see \cite{Rol94}.  In this scenario, unlike
either model described above, the GRBs will occur in the center of the
host.

Until recently, the most spectacular example of GRB energetics was
seen with GRB 971214 at $z = 3.418$ (\cite{kul98}): the implied
isotropic energy released from the burst in the $\gamma$-rays alone
was $E_\gamma \approx 3 \times 10^{53}$ erg, some two orders of
magnitude higher than the commonly assumed numbers.  However, this was
further surpassed by an order of magnitude by the recent discovery of
GRB 990123.

Following the detection by BeppoSAX (\cite{gcn199}), an optical
transient was discovered at Palomar (\cite{gcn201}), and subsequently
a coincident radio transient was found at the VLA (\cite{gcn211}),
within the error-circles of the GRB itself and the associated new
x-ray source (\cite{gcn203}).  Examination of the ROTSE images taken
within minutes of the burst revealed an unprecedented bright
($m_{peak} \approx 8.9$ mag) phase of the optical afterglow
(\cite{gcn205}).  Spectroscopy of the OT obtained at the Keck-II 10-m
telescope revealed an absorption system with $z_{abs} \approx 1.61$
(\cite{iauc7096}).  Together with the GRO/BATSE measurement of the
burst fluence (\cite{gcn224}), this implied a phenomenal energetics
for the burst and its afterglow.  The absorption redshift was
subsequently confirmed independently by \cite{gcn219}, and further
refinement of the spectroscopy improved the redshift measurement to
$z_{abs} = 1.6004$ (\cite{gcn249}, \cite{gcn251}).  A fading infrared
counterpart was discovered at the Keck-I 10-m telescope
(\cite{gcn240}).

The early reports indicated a presence of an apparent foreground
galaxy within $\sim 2$ arcsec from the OT (\cite{gcn201},
\cite{gcn206}, \cite{gcn213}) and, later found, the presence of
foreground emission and absorption lines at $z = 0.210$ and $z =
0.286$ (\cite{gcn219}).  Motivated by these reports, and the
unprecedented apparent energetics of the burst, it was proposed that
this burst may have been gravitationally lensed (\cite{gcn216}).
However, subsequent observations and analysis did not confirm the
existence of this foreground galaxy (\cite{gcn242}, \cite{gcn243}) nor
low-redshift absorbers close to the line-of-sight (\cite{gcn249},
\cite{gcn251}).  Thus, the empirical motivation for the gravitational
lensing of this burst was all but removed leaving open the problem of
its energetics.

\cite{Kul99} present a detailed study of the ground-based work on this
burst to date, and analyze its physical properties and energetics.
Early ground-based observations are dominated by the
afterglow light, which makes difficult the detection and study of the
host galaxy (and possible foreground objects near the line of sight).
In this {\it Letter} we report on the Hubble Space Telescope (HST)
observations of the host galaxy of this burst, about 16 days after the
burst itself, as well as the ground-based Keck imaging in the
near-infrared, starting from about 6 days after the burst.

\section{Observations and Data Reductions}

The ground-based near-IR images of the field were obtained using the
NIRC instrument (\cite{NIRC}) at the Keck-I 10-m telescope.  A log of
the observations and a detailed description of the data and the
reduction procedures are given by \cite{Kul99}.  The observations were
obtained in the $K$ or $K_s$ bands, and were calibrated to the
standard $K$ band ($\lambda_{eff} = 2.195 ~\mu$m).  The Galactic
extinction corrections are negligible in the $K$ band, assuming
$E_{B-V} = 0.016$ in this direction (\cite{SFD}).

The first evidence of the underlying galaxy, approximately 0.5 arcsec
from the OT was seen in the Keck images taken on 27 January 1999 UT.
We estimated a magnitude $K \sim 22$ mag (\cite{gcn243}).  The galaxy,
the putative host, which we designate as ``A'', was then clearly
detected in the images obtained on 29 January 1999 UT (\cite{gcn256}),
as shown in Figure 1.  The total $K$-band magnitude of the OT plus the
galaxy at that time (January 29.665 UT) was $K_{tot} = 20.30 \pm 0.10$
mag (including both random and systematic errors).  We estimate the
contribution of the galaxy to the total flux by masking the
appropriate pixels of the transient, and find that the galaxy
contributed about 21\% $\pm$ 5\% of the total $K$-band light at that
time, implying the magnitudes $K_{OT} = 20.56 \pm 0.17$ mag (at this
epoch), and $K_{gal.~A} = 22.0 \pm 0.4 $ mag.

\begin{figure}

\centerline{\psfig{file=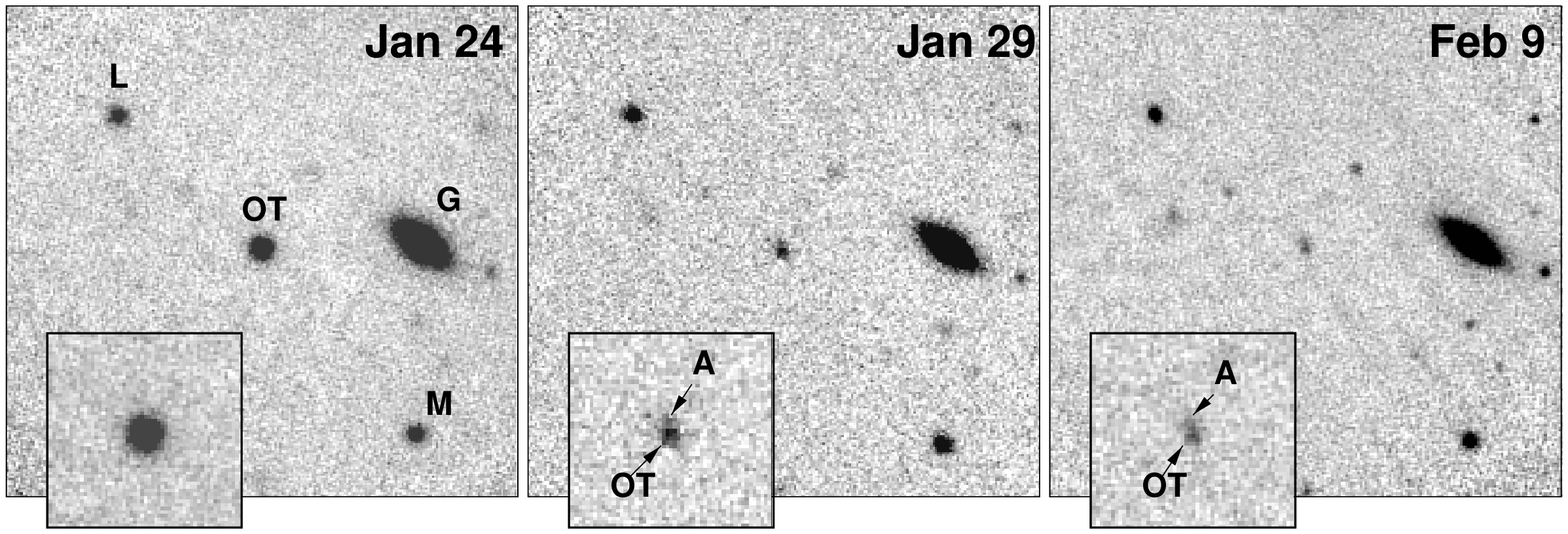,angle=-90,width=7.5in}}
\caption[]{Three epochs of Keck I $K$-band imaging of the field of GRB
990123 (24 January 1999 UT, 29 January 1999, and 9 February 1999 UT).
The field shown is 32 arcsec $\times$ 32 arcsec, corresponding to
about 270 physical kpc (710 comoving kpc) in projection at $z =
1.6004$ (for $H_0 = 65$ km s$^{-1}$ Mpc$^{-1}$ and $\Omega_0 = 0.2$).
The image is rotated to the standard orientation, so that the east is
to the left and north is up. In the 24 Jan~image, the OT dominates
the host galaxy flux, but by 29 Jan~the galaxy is resolved (see
inset) from the OT.}
\label{fig1}

\end{figure}

\placefigure{fig1}

The total $K$-band magnitude measured on 9 February 1999 UT is
$K_{tot} = 21.04 \pm 0.11$ mag.  If the OT had a power-law spectrum
$F_\nu \sim \nu^\beta$ with $\beta = -0.8$ and the
extinction-corrected Gunn $r$-band flux at the same epoch $F_r = 0.263
\pm 0.055 ~\mu$Jy (\cite{Kul99}), then the predicted $K$ band
magnitude of the OT alone would be $K_{OT} = 22.41 \pm 0.21$ mag (at
this epoch), and the resulting magnitude of the galaxy would be
$K_{gal.~A} = 21.40 \pm 0.20$ mag.  However, the slope of the OT
power-law continuum may have changed by that time, and if we assume
$\beta = -1.0$ instead, we derive $K_{OT} = 22.14 \pm 0.21$ mag, and
$K_{gal. A} = 21.53 \pm 0.23$ mag.  We thus assume the estimate of
$K_{gal. A} = 21.45 \pm 0.25$ mag from this decomposition.

Taking the weighted average of the two estimates, we find 
$K_{gal. A} = 21.65 \pm 0.30$ mag, corresponding to the flux
$F_{\nu,K,gal. A} = 1.39 \pm 0.44 ~\mu$Jy.  We assume for the flux
zero-point of the $K$ band 636 Jy for $K = 0$ mag (\cite{Bes88}).

The HST observations of the GRB 990123 field were obtained on 8--9
February 1999 UT in response to the Director's Discretionary time
proposal GO-8394, with the immediate data release to the general
community.  The CCD camera of the Space Telescope Imaging Spectrograph
(STIS) (\cite{Kimble}) in Clear Aperture (50CCD) mode was used.  The
CCD has a peak quantum efficiency at $\lambda = 5852$ \AA\ over the
wavelength range $\lambda 2000 - 10000$ \AA.  A total of 6 exposure of
1300 sec each was collected over 3 orbits.  The field was imaged in
six positions dithered in a spiral pattern. Each position was imaged
twice to facilitate cosmic-ray removal (total of 12 integrations).

Initial data processing followed the STScI pipeline procedures,
including bias and dark current subtraction.  The six cosmic-ray
removed images were then combined by registering 
the images and median stacking to produce a master science-grade image.  
Photometry and astrometry were performed with this image. 

Figure 2 shows a portion of the STIS image of the GRB 990123 field.  
We find (see below) the OT as a point source clearly detected
$0.648 \pm 0.1$ arcsec to the south of galaxy A. Galaxy A has an
elongated and clumpy appearance, possibly indicative of star formation
regions in a nearly edge-on (potentially late-type) disk galaxy,
although a classification as a purely Irregular galaxy cannot be
excluded.  Such morphologies are typical for many galaxies at
comparable flux levels, as observed with the HST.  Its extension to
the south clearly overlaps with the OT, and it is thus virtually
certain that this galaxy is responsible for the absorption line system
at $z_{abs} = 1.6004$ (\cite{iauc7096}, \cite{Kul99}).

\begin{figure}
\centerline{\psfig{file=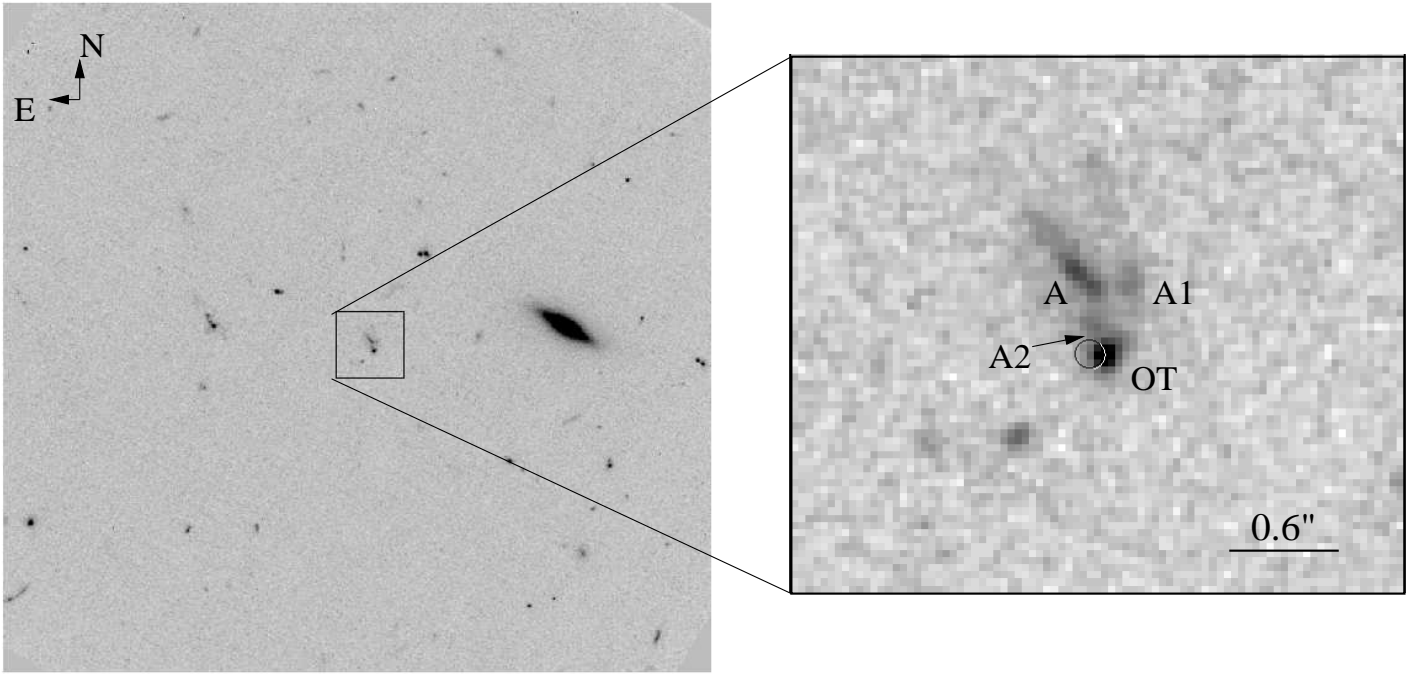,angle=-90,width=7.5in}}
\caption[]{(left) The HST STIS image of the field of GRB 990123
rotated to the normal orientation.  The field shown is $38 \times 38$
arcsec square, corresponding to about 330 $\times$ 330 physical kpc
square (850 $\times$ 850 comoving kpc square) in projection at $z =
1.6004$ (for $H_0 = 65$ km s$^{-1}$ Mpc$^{-1}$ and $\Omega_0 = 0.2$).
(right) The optical transient (OT), galaxy A (the putative host), and
the bright knot (A1) associated with galaxy A are denoted.  The
ellipse which overlays the OT point-source is the 1-$\sigma$
uncertainty contour for the position of the OT as measured in
ground-based imaging (see text).  The positional consistency
definitely establishes that the point source is indeed the OT. A small
nebulousity (A2) just to the north of the OT may be indicative of a
star-forming region.}
\label{fig2}
\end{figure}

\placefigure{fig2}

  Our earlier ground-based imaging suggested the OT was displaced to the 
south of galaxy A, and we now confirm this with a precise astrometric 
tie between the discovery image and the STIS image.  We measured the 
centroid of the optical transient in our discovery image from Jan 23 
(\cite{gcn201}) at the
Palomar 60-inch (P60). The OT was bright ($r$ = 18.65) at this early
epoch and its position is well-determined with respect to other objects
in the field.  Next we computed the astrometric mapping of the P60
coordinate system to a deep Keck II 10-m R-band image (see
\cite{Kul99}) using 75 well-centroided objects common to the two
images. Similarly, we tie the Keck II coordinates to the STIS image
using 19 common tie objects.  We find the ground-based
position of the OT is consistent with the STIS point-source with a
negligible offset of $0.09 \pm 0.18$ arcsec.

  The coordinates of the OT as measured in the HST image are
\hbox{$\alpha = $ 15\fh 25\fm 30\fs .3026}, \hbox{$\delta = +44\fd
45\arcmin 59\arcsec .048$} (J2000). This is in an excellent agreement
with an absolute astrometric measurement, tied to the USNO A-2.0 
catalogue (\cite{monet98}), from the ground-based image discussed in  
\cite{gcn206}. The coordinates of the center of galaxy ``A'' 
(extended North-South) are \hbox{$\alpha = $ 15\fh 25\fm 30\fs .3175}, 
\hbox{$\delta = +44\fd 45\arcmin 59\arcsec .676$} (J2000). 
The brightest knot connected with
galaxy A is located at \hbox{$\alpha = $ 15\fh 25\fm 30\fs .2835},
\hbox{$\delta = +44\fd 45\arcmin 59\arcsec .576$} (J2000).  The
uncertainties in the relative positions are $\sim 10$ mas, but the
positions in an absolute sense (relative to Hipparcos) are larger
($\sigma \simeq 0.3$ arcsec).

No other galaxies brighter than $V \sim 27$ mag are detected in the
STIS image closer to the OT than galaxy A, and we see no evidence for
a distant cluster (or even a sizable group) in this field.  This
effectively removes the possibility that the burst was significantly
magnified by gravitational lensing.

We will assume for the Galactic reddening in this direction $E_{B-V} =
0.016$ mag (\cite{SFD}), and use the standard Galactic extinction
curve with $R = A_V/E_{B-V} = 3.1$ to estimate extinction corrections
at other wavelengths.  We assume the photometric flux zero points as
tabulated by \cite{Fukugita}.

In order to convert the observed counts to fluxes and magnitudes, we
use the web-based STIS exposure simulator (ETC).  Since the bandpass
of the STIS CLEAR is so broad, the conversion depends on the assumed
spectrum of the object.  For the OT, we can assume a power-law
spectrum $F_\nu \sim \nu^\beta$ with $\beta = -0.8$ (\cite{Kul99}).
At the effective wavelength of the STIS bandpass, $\lambda_{eff}
\approx 5850$ \AA, for the OT alone we derive $F_{\nu, OT} = 0.308
~\mu$Jy, uncertain by a few percent.  Applying the Galactic extinction
correction increases that number by about 5\%.  Assuming the same
power-law, we derive the extinction-corrected magnitudes $V_{OT} =
25.17$ mag in the standard Johnson system, and $r_{OT} = 24.82$ mag in
the Gunn-Thuan system.

For galaxy A, using the same power-law spectrum (which is a good
approximation for star-forming galaxies with modest extinction, in
this redshift range), we obtain $F_{\nu, gal. A} = 0.648 ~\mu$Jy,
uncertain by a few percent, at $\lambda_{eff} \approx 5850$ \AA,
before the extinction correction.  If we assume $\beta = 0$ instead
(as it may be appropriate in the rest-frame UV for an actively
star-forming galaxy and no extinction), we obtain $F_{\nu, gal. A} =
0.615 ~\mu$Jy.  The difference is indicative of the net uncertainty of
the flux conversion.  We thus derive for the galaxy alone, in the
$\beta = -0.8$ case: $V_{OT} = 24.36$ mag, $r_{OT} = 24.01$ mag, and
the flux at $\lambda_{obs} = 7280$ \AA, corresponding to
$\lambda_{rest} = 2800$ \AA, $F_{\nu, 2800} = 0.798 ~\mu$Jy (all
corrected for the Galactic extinction).  If we assume $\beta = 0$,
these values become: $V_{OT} = 24.37$ mag, $r_{OT} = 24.17$ mag, and
$F_{\nu, 2800} = 0.637 ~\mu$Jy.

We note that the simple power-law approximation to the broad-band
spectrum of the galaxy, as defined by our STIS and $K$-band
measurements, is $\beta_{gal. A} = -0.65 \pm 0.1$.  This relatively
blue color is suggestive of active star formation, but it cannot be
used to estimate the SFR directly.

\section{Discussion}

We will assume a standard Friedman model cosmology with $H_0 = 65$ km
s$^{-1}$ Mpc$^{-1}$, $\Omega_0 = 0.2$, and $\Lambda_0 = 0$.  For $z =
1.6004$, the luminosity distance is $3.7 \times 10^{28}$ cm, and 1
arcsec corresponds to 8.64 proper kpc or 22.45 comoving kpc in
projection.

It is practically certain that the absorption system at $z_{abs} =
1.6004$ originates from galaxy A, as no other viable candidate is seen
in the HST images.  The proximity of the center of galaxy A to the OT
line of sight ($0.638 \pm 0.1$ arcsec), corresponding to 5.5 proper
kpc at this redshift, strongly suggests that the two are physically
related, and we propose that A is the host galaxy of the GRB.  Visual
inspection of Figure 2 suggests that a probability of chance
superposition on this magnitude level is negligibly small.

In order to estimate the rest-frame luminosity of galaxy A, we
interpolate between the observed STIS and $K$-band data points using a
power-law, to estimate the observed flux at $\lambda_{obs} \approx
11570$ \AA, corresponding approximately to the effective wavelength of
the rest-frame $B$ band.  We obtain $F_{\nu, B, rest} \approx 1.03
~\mu$Jy, corresponding to the absolute magnitude $M_B = -20.4$.  At $z
\sim 0$, assuming $H_0 = 65$ km s$^{-1}$ Mpc$^{-1}$, an $L_*$ galaxy
has $M_B \approx -20.75$.  We thus conclude that this object has the
rest-frame luminosity $L \approx 0.7 ~L_{*,now}$.  Given the
uncertainty of the possible evolutionary histories, it may evolve to
become either a normal spiral galaxy, or a borderline dwarf galaxy.

We can make a rough estimate of the SFR from the continuum luminosity
at $\lambda_{rest} = 2800$ \AA, following \cite{MAD98}.  Using the
$F_{\nu, 2800}$ estimates given above, the corresponding monochromatic
rest-frame power is $P_{\nu, 2800} = 4.21 \times 10^{28}$ erg s$^{-1}$
Hz$^{-1}$ (for $\beta = 0$, as it may be appropriate in the UV
continuum itself), or $P_{\nu, 2800} = 5.28 \times 10^{28}$ erg
s$^{-1}$ Hz$^{-1}$ (for $\beta = -0.8$).  The corresponding estimated
unobscured star formation rates are $SFR \approx 5.3~M_\odot$
yr$^{-1}$ and $SFR \approx 6.7~M_\odot$ yr$^{-1}$, probably accurate
to within 50\% or better.  This is a relatively modest value, but it
may be typical for normal galaxies at such redshifts.  It is of course
a lower limit, as it does not include any extinction corrections in
the galaxy itself, or any fully obscured star formation.

Further insight into the physical properties of this galaxy comes from
its absorption spectrum, presented in \cite{Kul99}.  The lines are
unusually strong, placing this absorber in the top 10\% of all Mg II
absorbers detected in complete surveys (\cite{steidel}).
Unfortunately, without a direct measurement of the hydrogen column
density, it is impossible to estimate the metallicity of the gas.
However, the data suggest that the gas originated in a chemically
evolved, massive galaxy.  We note that strong metallic line absorbers
are frequently associated with high hydrogen column density systems,
such as damped Ly$\alpha$ absorbers.  The scatter in the redshift
measurements of the individual lines implies a very small velocity
dispersion, less than 60 km s$^{-1}$ in the galaxy's rest-frame
(\cite{gcn251}), implying that the absorber is associated either with
a dwarf galaxy, or a dynamically cold disk of a more massive system, a
possibility which we consider to be more likely.

It is worth examining the observed offset between the OT and the
galaxy's center, in the context of previously studied cases.  At least
five GRBs now have offsets between the centroid of the visible host
and the OT measured with sufficient accuracy to test association with
galactic nuclei.  Figure 3 shows the measured offset between the
centroid of the host galaxy and the OT as a function of the host
galaxy magnitude.  Two of these offsets are based on previous STIS
observations with HST: GRB 970228 by \cite{Fruch_0228} and GRB 971214
by \cite{Ode98}. The host magnitude is correlated with galaxy radius,
and we use it as an objective measure of the host size.  The two
curves in Fig.~3 represent the median trend in effective radius
($\sim$ half-light radius) as a function of apparent magnitude.
Smooth relations were computed with overlapping magnitude bins using
approximately 1304 faint galaxies measured in F814W and F606W in two
deep WFPC2 fields by \cite{Ode96}. Applying these mean trends to the
total magnitude of all known host galaxies to date we note that the
optical transients (except for 990123) lie well within the effective
radius predicted for each host.

While the offset of the OT with respect to the effective radius (as
inferred from observed magnitude) is a useful estimate of the relation
of the GRB to the host, resolved imaging using HST provides the
clearest picture. The transient of GRB 970228 is clearly displaced
from its host center, but within the effective radius.  GRB 970508, on
the other hand, is coincident with the nucleus of its host galaxy to
0.01 arcseconds.  Lastly, GRB 990123 is clearly separated from the
central region of host and appears to be coincident with a
star-forming region (see Fig.~2).

It has been clear since the discovery of GRB optical transients that
GRBs are connected with galaxies. Gradually, however, a more specific
picture has emerged.  Imaging of the transient of GRB 990123 and its
host is perhaps the best direct observational evidence that GRBs are
intimately connected with the formation of stars: GRBs are clearly not
a nuclear phenomenon, nor do most occur far-outside their host. The
present suggestion of the spatial coincidence of GRB 990123 with a
star-formation regions opens the possibility of studying not just GRBs
in galaxies, but GRBs in their host environments.

\begin{figure}
\figurenum{3}

\centerline{\psfig{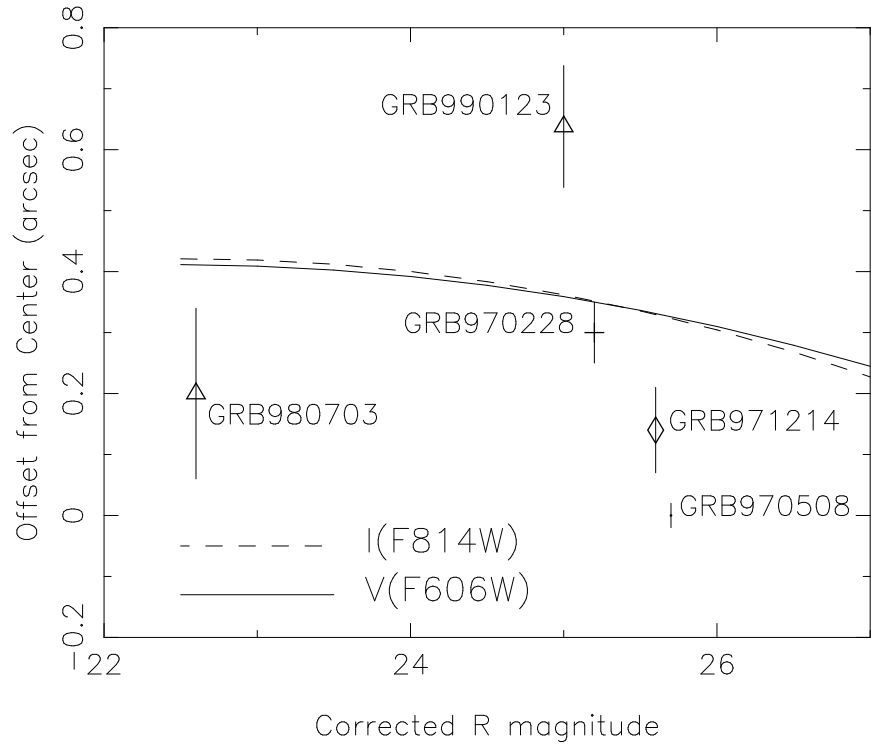}}
\caption{Projected separations of OTs from the centers of their host
galaxies, as a function of the host galaxy's $R$ band
magnitude. Except for GRB 990123, the measured offset of the OT falls
within the predicted effective radius (solid and dashed curves) of the
host galaxy. Clearly, most GRBs cannot be associated with nuclear
activity (e.g.~AGN), nor do most GRB events occur far from stellar
birth sites within the host galaxy. }
\label{fig3}

\end{figure}

\placefigure{fig3}

\acknowledgments

We are grateful to S.~Beckwith of STScI for the allocation of the
Director's Discretionary time for this project, and to the entire
BeppoSAX team and the staff of W.~M.~Keck Observatory for their
efforts.  This work was supported in part by the HST grant GO-8394,
the grants from the NSF and NASA (SRK, JSB), the Bressler Foundation
(SGD, SCO, RRG).

\clearpage

\clearpage

\end{document}